\documentclass[10pt,twocolumn,letterpaper]{article}

\usepackage{wacv}
\usepackage{times}
\usepackage{epsfig}
\usepackage{graphicx}
\usepackage{amsmath}
\usepackage{amssymb}
\usepackage{booktabs}

\usepackage{graphicx}
\usepackage{multirow}
\usepackage{booktabs}
\usepackage{caption} 
\usepackage{breqn}

\usepackage[accsupp]{axessibility} 

\def\wacvPaperID{1372} 

\wacvapplicationstrack 

\wacvfinalcopy 

\ifwacvfinal
\usepackage[breaklinks=true,bookmarks=false]{hyperref}
\else
\usepackage[pagebackref=true,breaklinks=true,colorlinks,bookmarks=false]{hyperref}
\fi

\begin{document}

\title{Self-Supervised 2D/3D Registration for X-Ray to CT Image Fusion}

\author{{Srikrishna Jaganathan$^{1,2}$} ~ {Maximilian Kukla$^2$}  ~ {Jian Wang$^2$} ~ {Karthik Shetty$^1$} ~ {Andreas Maier$^1$} \\
\normalsize $^1$FAU~Erlangen-Nürnberg, Erlangen, Germany \quad $^2$Siemens~Healthineers~AG, Forchheim, Germany\\
{\tt\small srikrishna.jaganathan@fau.de}
}

\maketitle
\thispagestyle{empty}

\begin{abstract}

Deep Learning-based 2D/3D registration enables fast, robust, and accurate X-ray to CT image fusion when large annotated paired datasets are available for training. However, the need for paired CT volume and X-ray images with ground truth registration limits the applicability in interventional scenarios. An alternative is to use simulated X-ray projections from CT volumes, thus removing the need for paired annotated datasets. Deep Neural Networks trained exclusively on simulated X-ray projections can perform significantly worse on real X-ray images due to the domain gap. We propose a self-supervised 2D/3D registration framework combining simulated training with unsupervised feature and pixel space domain adaptation to overcome the domain gap and eliminate the need for paired annotated datasets. Our framework achieves a registration accuracy of $1.83 \pm 1.16$ mm with a high success ratio of 90.1\% on real X-ray images showing a 23.9\%  increase in success ratio compared to reference annotation-free algorithms.

\end{abstract}

\section{Introduction}

Image guidance for minimally invasive interventions is generally provided using live fluoroscopic X-ray imaging. The fusion of preoperative Computed Tomography (CT) volume with the live fluoroscopic image enhances the information available during the intervention. Spatial alignment of the 3D volume on the current patient position is a prerequisite for accurate fusion with the fluoroscopic image. An optimal spatial alignment between preoperative CT volume and live fluoroscopic X-ray is estimated with 2D/3D registration. Traditionally, optimization-based techniques have been used for 2D/3D registration in the interventional setting as it provides highly accurate registration~\cite{Wang2017, Markelj2012, wang2020robust}. However, optimization-based techniques are sensitive to initialization and content mismatch between X-ray and CT images. Deep Learning (DL)-based 2D/3D registration techniques have been proposed to overcome the limitations of the optimization-based techniques by improving the robustness significantly~\cite{Liao2020, Miao2018, Miao2016a, Schafferta}, while still relying on optimization-based techniques as a subsequent refinement step to match the registration accuracy. Recently, end-to-end DL-driven solutions have been proposed that can achieve a combination of high registration accuracy and high robustness with faster computation~\cite{Jaganathan2021DeepRegistration}. 

Despite the significant improvement in learning-based registration techniques, the interventional application is still limited due to the lack of generalizability of the learned networks for different anatomy, interventions, scanner, and protocol variations~\cite{unberath2021impact}. The collection of large-scale annotated datasets for all variations is prohibitive since the data needed for training should be paired along with ground truth registration. Either a large-scale annotated dataset that consists of all the different variations or an annotation-free unpaired training routine based on existing DL-based technique enables us one step closer to interventional application. We focus on the latter, by removing the need for annotated paired dataset as this would immediately allow us to train the current state-of-the-art registration networks for different variations.

We propose a self-supervised 2D/3D rigid registration framework to achieve annotation-free unpaired training with minimal performance drop on real X-ray images encountered during the interventional application.
The annotation-free unpaired dataset is generated from forward projections of the CT volumes. Our framework consists of simulated training combined with unsupervised feature and pixel space domain adaptation. Our novel task-specific feature space domain adaptation is trained in an end-to-end manner with the registration network. We combine the recently proposed Barlow Twins~\cite{ZbontarBarlowReduction}, adversarial feature discriminator~\cite{Ganin2015Domain-AdversarialNetworks, Karaoglu2021AdversarialEstimation} and DL-based registration network~\cite{Jaganathan2021DeepRegistration}. This allows the features to be robust for different style variations while also being optimal for the registration task. Our feature space adaptation adds no computational cost during inference. We additionally perform unsupervised style transfer of the real X-ray to simulated X-ray image style using Contrastive Unpaired Translation~\cite{Park2020ContrastiveTranslation}. We apply the style transfer network during inference, thus allowing the registration network to operate on the fixed style already encountered during training.
In combination, our proposed framework achieves a registration accuracy of \mbox{$1.83 \pm 1.16$ mm} with a high success ratio of 90.1\% on real X-ray images showing a 23.9\% increase compared to reference annotation-free techniques.

\section{Related Work}

We focus our related work discussion specific to rigid 2D/3D registration for optimization-based and learning-based 2D/3D registration algorithms. In unsupervised domain adaptation, we broadly discuss the methods applied in medical imaging tasks.

\paragraph{Optimization-Based 2D/3D Registration}
The problem of 2D/3D registration for interventional image fusion has been extensively researched with comprehensive reviews of the techniques available~\cite{liao2013review, Markelj2012}. Due to the non-convex nature of the 2D/3D registration problem, global optimization~\cite{gong20082d, grupp2020automatic, otake2013robust} is required to reach optimal solution. However, the high computational cost of global optimization-based techniques limits the interventional application. Faster techniques using local optimization-based methods~\cite{tomazevic20033,mitrovic2011gradient, Matl2017VascularReview, groher20082d, vspiclin2014fast} rely on image similarity measures, making it highly dependent on good initialization. Point-to-Plane Correspondence (PPC) constraint was proposed~\cite{Wang2017, wang2014gradient, wang2020robust} as a more robust alternative for computing the 3D motion from the 2D misalignment visible between the 2D image and the forward projection of the 3D volume. PPC-based techniques significantly improve the registration accuracy and robustness compared to other optimization-based techniques. Extensions of the \mbox{PPC-based} technique proposed for multi-view scenario~\cite{schaffert2019robust} and hybrid learning-based solutions improve the robustness significantly~\cite{Schafferta}.
Recently, multi-level optimization-based technique~\cite{lange2020multilevel} was proposed with normalized gradient field as the image similarity metric, showing further improvement in the registration accuracy.

\paragraph{Learning-Based 2D/3D Registration}
Initially, learning-based techniques were targeted to improve the computational efficiency~\cite{Miao2016a} and robustness~\cite{Miao2018, Liao2020, Schafferta, gu2020extended, gao2020generalizing} of the \mbox{optimization-based} techniques. DL-based techniques significantly improve the robustness to initialization and content mismatch~\cite{Schafferta, Liao2020, Miao2018}. End-to-end DL-driven registration~\cite{Jaganathan2021DeepRegistration} has shown improved robustness compared to other learning-based methods~\cite{Schafferta,jaganathan2021learning, schaffert2020learning}, while also matching the registration accuracy of the \mbox{optimization-based} techniques~\cite{wang2020robust} with significant improvement in computational efficiency. Recently, fully automatic DL-based registration has been proposed ~\cite{esteban2019towards, grupp2020automatic, Grimm2021} that can perform both initialization and registration. A comprehensive review of the learning-based medical image registration~\cite{Haskins2020DeepSurvey} and the impact of learning-based 2D/3D registration for interventional applications~\cite{unberath2021impact} are available. The advances in DL-based 2D/3D registration techniques have been propelled by using supervised techniques~\cite{Jaganathan2021DeepRegistration, Miao2018, Liao2020}. The variations in imaging protocol, device manufacturer, anatomy, and intervention-specific setting alter the appearance of the acquired images significantly, preventing the adoption of the DL-based 2D/3D registration techniques in interventional scenarios. Attempts have been made to reduce the number of annotated data samples required with paired domain adaptation techniques~\cite{zheng2017learning, zheng2018pairwise}. 
Simulated X-ray projections generated from CT volume remove the need for paired annotated data requirement. However, this leads to a domain gap due to the variations between the real and simulated X-ray projection. Unsupervised domain adaptation is required to minimize the drop in performance while not requiring annotated datasets.

\paragraph{Unsupervised Domain Adaptation}
The domain gap introduced due to the use of simulated data is bridged either by improving the realism of the simulated data~\cite{Wood2021FakeAlone, Unberath2018} or by performing unsupervised domain adaptation to minimize the domain gap~\cite{Armanious2018MedGAN:GANs, Ganin2015Domain-AdversarialNetworks, Zhu2017UnpairedNetworks, Park2020ContrastiveTranslation,Venator2020Dual-modeAdaptation, hoffman2018cycada, tobin2017domain, Hoffmann2020SynthMorph:Images}. The use of simulated X-ray projections is increasing in training DL-based solutions~\cite{Grimm2021, Bier2018X-ray-transformSurgery, Zhang2018TaskSegmentation, ying2019x2ct, zheng2021unsupervised, sukesh2022training} for various medical imaging applications.
DeepDRR~\cite{Unberath2018}, aims to bridge the domain gap by rendering realistic simulated X-ray projections which are closer to real X-ray images. 
Domain Randomization~\cite{tobin2017domain} was recently proposed to bridge the domain gap problem in learning-based 2D/3D registration networks~\cite{Grimm2021}. Multiple different styles of simulated X-ray images  are used during training, allowing the network to be robust to style variations encountered during inference. However, a patient-specific retraining step is required on top of domain randomization~\cite{Grimm2021}. 
Unsupervised domain adaptation for CNN-based 6D pose regression of X-ray images was proposed in~\cite{zheng2021unsupervised}. However, the evaluation ignores the use of surgical tools and content mismatch which is crucial for the interventional application.
Generative Adversarial Network (GAN)~\cite{Armanious2018MedGAN:GANs, Zhang2018TaskSegmentation, ying2019x2ct, zheng2021unsupervised} and 
adversarial feature adaptation techniques~\cite{Karaoglu2021AdversarialEstimation, dou2018pnp,zheng2021unsupervised, vesal2021adapt} have shown promising results for bridging the domain gap in medical imaging tasks like segmentation~\cite{Zhang2018TaskSegmentation}, reconstruction~\cite{ying2019x2ct}, pose regression~\cite{zheng2021unsupervised}, depth estimation~\cite{Karaoglu2021AdversarialEstimation} and multi-modality learning~\cite{vesal2021adapt, dou2018pnp}.

\section{Methods}

\begin{figure}
    \centering
    \includegraphics[width=\linewidth]{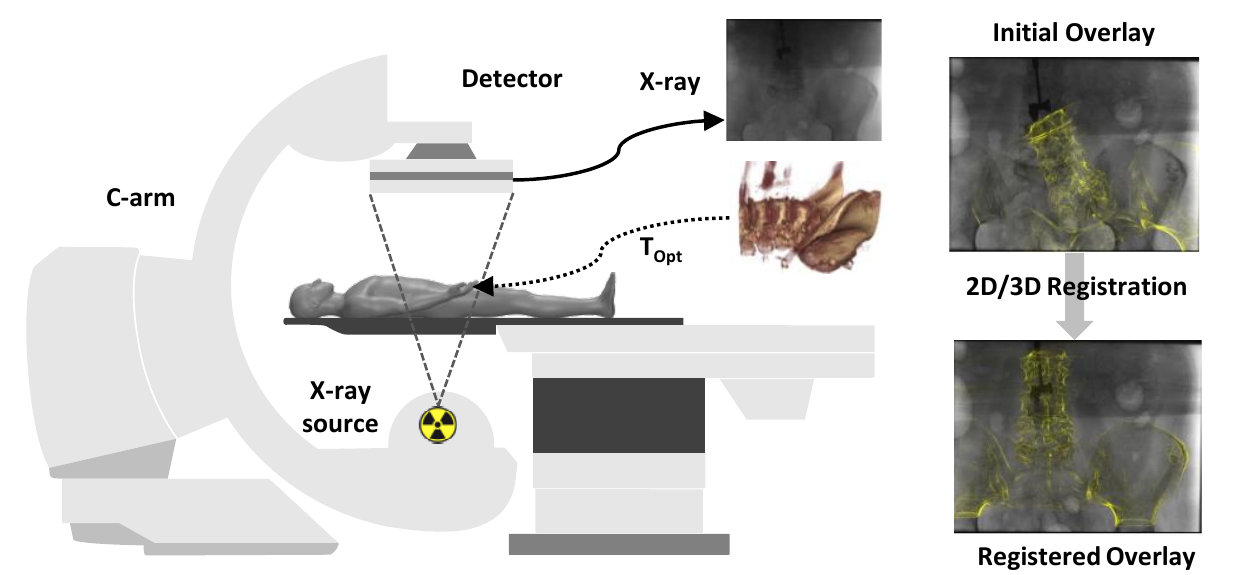}
    \caption{Interventional image fusion for C-arm system, showing the overlay before and after performing 2D/3D registration.}
    \label{fig:reg_intro}
\end{figure}

Our proposed method is targeted towards rigid 2D/3D registration for interventional image fusion between X-ray (2D) and CT (3D) images as illustrated in Figure~\ref{fig:reg_intro}. The live X-ray image is acquired from the C-arm system during the intervention. The initial overlay depicts the fusion of the 3D volume with the 2D image after performing initialization (either manually or automatically). The registered overlay depicts the overlay produced after performing 2D/3D registration which spatially transforms the preoperative volume with the patient's position and orientation. 
We give a brief introduction to the 2D/3D registration problem and the registration framework on top of which we build our self-supervised method in Section~\ref{sec:background}. Following, we describe our proposed self-supervised registration technique, with the different components that are used during training and inference in Section~\ref{sec:ss_dirn}. We finally describe the training and inference procedure used for our registration framework in Section~\ref{sec:training}.

\subsection{Background}
\label{sec:background}

\paragraph{Problem Formulation}

In 2D/3D registration, the goal is to find an optimal spatial transformation $\mathbf{T}_{\mathrm{opt}}$ of the volume $\mathbf{V}$ from the observed X-ray projections $\mathbf{I}_{f}^r$ such that when the images are overlaid, there is minimal misalignment. The problem can be formulated as an optimization problem with an objective function $\mathcal{F}$ that minimizes the misalignment as described in Eq.~\ref{eq:opt_problem}.  The X-ray projection $\mathbf{I}_{f}^r$, the preoperative volume $\mathbf{V}$ and an initial registration estimate $\mathbf{T}_{\mathrm{init}}$ are given as input to the registration algorithm. Our focus is on recovering the registration matrix $\mathbf{T}_{\mathrm{reg}}$ which  enables us to find the optimal transformation $\mathbf{T}_{\mathrm{opt}} = \mathbf{T}_{\mathrm{reg}} \mathbf{T}_{\mathrm{init}}$ that aligns the forward projected 3D volume $\mathcal{R}(\mathbf{V},\mathbf{T})$ with the X-ray projection $\mathbf{I}^r_f$.

\begin{equation}
  \operatorname*{argmin}_{\mathbf{T}} \ \mathcal{F} (\mathbf{I}_{f}^r, \mathcal{R}(\mathbf{V}, \mathbf{T}))
  \label{eq:opt_problem}
\end{equation}

\paragraph{Forward Projection}

\begin{figure}
    \centering
    \includegraphics[width=\linewidth]{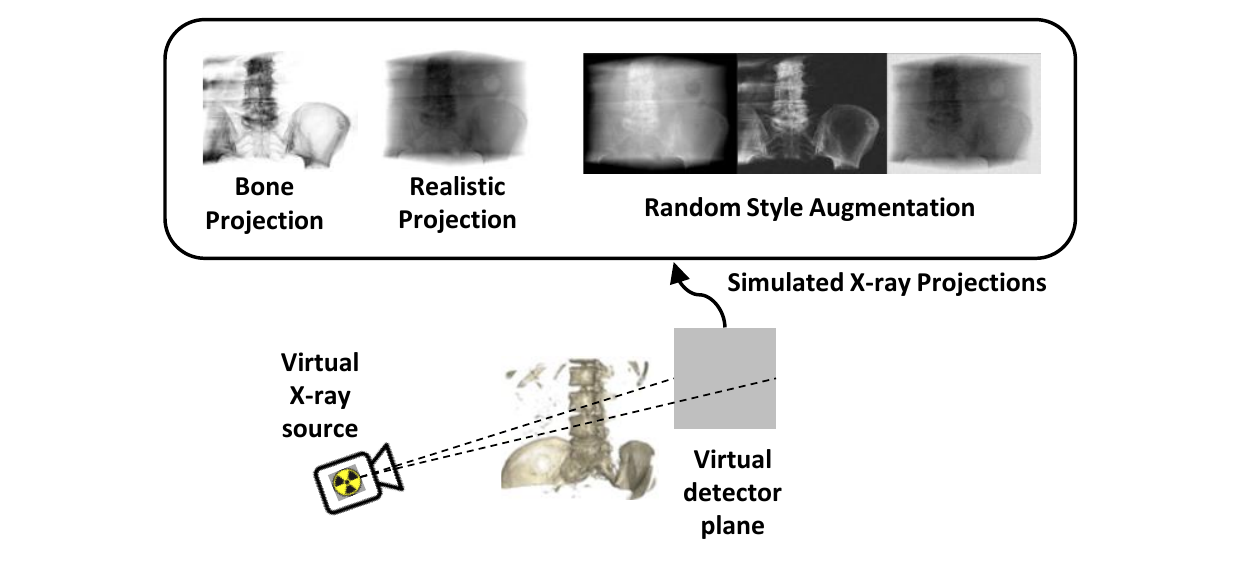}
    \caption{Rendering simulated X-ray projection from 3D CT volume with different styles.}
    \label{fig:sim_xproj}
\end{figure}

A common basis for comparison between the 2D and 3D images is established using the forward projector (rendering) $\mathcal{R}(\mathbf{V}, \mathbf{T})$, which is used to generate simulated X-ray projection $\mathbf{I}_{m}$ also referred to as Digitally Reconstructed Radiograph (DRR).
The forward projection from a CT volume to render a DRR is depicted in Figure~\ref{fig:sim_xproj}. The rendering is performed by computing each detector pixel's attenuation response from a virtual X-ray source analytically using ray tracing~\cite{Unberath2018}. The rendering can be done using GPUs allowing for real-time computation~\cite{birkfellner2005fast}. 
Arbitrary DRRs can be rendered from CT volume for different viewing angles by altering the position and orientation $\mathbf{T}$ of the virtual X-ray source and detector. The style of the rendering can be controlled by selecting the desired materials to be rendered from CT volume either using segmentation or a simple threshold. We show examples of bone projection and realistic projection styles in Figure~\ref{fig:sim_xproj}.  The bone projection uses thresholding to render only the bones from the CT volume. The realistic projection renders all the materials present in the CT volume. Additionally, random style augmentations (Figure~\ref{fig:sim_xproj}) of the projected DRR are obtained by adjusting contrast, brightness, inverting, and adding noise.

\paragraph{PPC-Based Registration Framework}

Point-to-Plane Correspondence (PPC)-based registration framework~\cite{Wang2017, wang2014gradient, wang2020robust} constraints the global 3D motion $\mathbf{dv}$  from the visible 2D misalignment using the PPC constraint described in Eq.~\ref{eq:ppc_constraint}. The framework requires as input, the 3D volume $\mathbf{V}$, X-ray image $\mathbf{I}^r_f$ and initial registration estimate $\mathbf{T}_{init}$. The contour points $\mathbf{w}$ and their gradients $\mathbf{g}$ are computed from $\mathbf{V}$ using a 3D canny edge detector~\cite{Wang2017}.
The 3D motion $\mathbf{dv}$ is estimated by solving the PPC constraint (Eq.~\ref{eq:ppc_constraint}). The motion estimation $\mathbf{dv}$ is applied iteratively until convergence~\cite{Wang2017}.
During each motion estimation step, the previous registration estimate $\mathbf{T}_{i-1}$ is used for rendering the DRR $\mathbf{I}_{m} =\mathcal{R}(\mathbf{V},\mathbf{T}_{i-1})$  with $\mathbf{T}_{0} = \mathbf{T}_{init}$. Correspondence is estimated for a set of projected contour points $\mathbf{p}$ between $\mathbf{I}_{m}$ and $\mathbf{I}_{f}^r$. The corresponding projected contour point $\mathbf{p}'$ in $\mathbf{I}_{f}^r$ is used to compute the 2D misalignment $\mathbf{dp} = \mathbf{p}'-\mathbf{p}$. The plane normal $\mathbf{n}$ in Eq.~\ref{eq:ppc_constraint} can be computed from $\mathbf{w}$, $\mathbf{g}$ and $\mathbf{dp}$.

\begin{dmath}
    \mathbf{W}~\underbrace{[\mathbf{n} \times \mathbf{w} , -\mathbf{n}]}_{\textrm{\mathbf{A}}}~\mathbf{dv} =  \mathrm{diag}(\mathbf{W}) ~\underbrace{\mathbf{n}^T \mathbf{w}}_{\textrm{\mathbf{b}}}
    \label{eq:ppc_constraint}
\end{dmath}

The 3D motion $\mathbf{dv}$ is in axis-angle representation and can be directly converted to a 3D transformation matrix $\mathbf{T}_{i}$ serving as the current registration estimate. To account for noisy correspondence, per correspondence weight (diagonal)  matrix  $\mathbf{W}$ is used in Eq.~\ref{eq:ppc_constraint}.

\paragraph{Deep Iterative 2D/3D Registration}
\begin{figure}
    \centering
    \includegraphics[width=\linewidth]{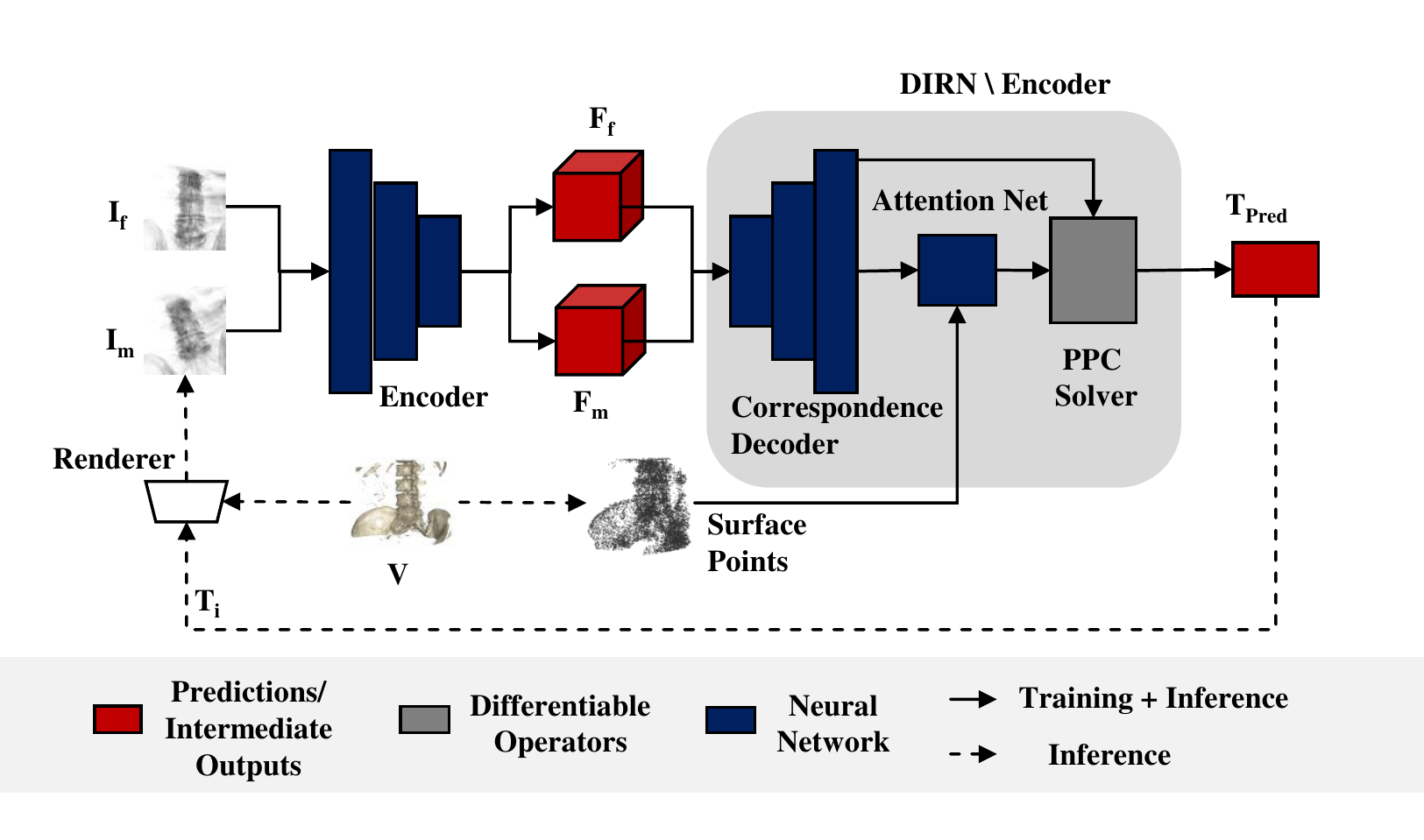}
    \caption{Schematic of DIRN with simulated training using DRRs generated from arbitrary views of the CT volume. The real X-ray image $\mathbf{I}_f^r$ is replaced with bone projection style DRR $\mathbf{I}_f$ for simulated training.}
    \label{fig:dirn}
\end{figure}
The closed form solution of Eq.~\ref{eq:ppc_constraint} is differentiable~\cite{Schafferta}, allowing the PPC constraint to be embedded as a known operator~\cite{maier2019learning} in \mbox{learning-based} 2D/3D registration methods~\cite{Schafferta, Jaganathan2021DeepRegistration, schaffert2020learning, jaganathan2021learning}.
We use the recently proposed Deep Iterative 2D/3D Registration Network (DIRN)~\cite{Jaganathan2021DeepRegistration} as the base architecture for our DL-based registration. 
Figure~\ref{fig:dirn} depicts the schematic of the DIRN-based registration framework.
In DIRN, the correspondence search and correspondence weighting of the classical PPC-based registration framework~\cite{Wang2017} is replaced by learned components.  RAFT~\cite{Teed2020RAFT:Flow} architecture (encoder and correspondence decoder in Figure~\ref{fig:dirn}) is used for estimating the correspondence between the  $\mathbf{I}_{f}$ and  $\mathbf{I}_{m}$ images.  The per correspondence weights are learned using a PointNet++~\cite{qi2017pointnet++} architecture (attention net in Figure~\ref{fig:dirn}) which takes $\mathbf{w}, \mathbf{g},\mathbf{p}', \mathbf{n}$. The predicted correspondences along with the predicted weights are used as inputs to the PPC solver. The 3D motion $\mathbf{dv}$ is computed using the closed form solution of Eq.~\ref{eq:ppc_constraint}. The network is trained for a single registration update using Eq.~\ref{eq:DIRN}. During inference, iterative application of the learned network for a fixed number of iteration is used for registration.

\begin{dmath}
    \mathcal{L}_{dirn} =  \mathcal{L}_{\mathrm{reg}} +  w_{\mathrm{flow}} \mathcal{L}_{\mathrm{flow}}  + w_{\mathrm{m}} \mathcal{L}_{m}  
    \label{eq:DIRN}
\end{dmath}
$\mathcal{L}_{\mathrm{reg}} = || \mathbf{T}(\mathbf{w}) - \mathbf{\hat{T}}(\mathbf{w}) ||$ is the registration loss with predicted transformation $\mathbf{T}$  computed from the predicted motion $\mathbf{dv}$, ground truth transformation $\mathbf{\hat{T}}$ and  3D contour points $\mathbf{w}$. $\mathcal{L}_{\mathrm{flow}}$ is the flow loss~\cite{jaganathan2021learning, Jaganathan2021DeepRegistration} and $\mathcal{L}_{m} = ||\mathbf{dv}||^2 $ is the regularization loss on the predicted 3D motion $\mathbf{dv}$. $w_{\mathrm{flow}}$, $w_{\mathrm{m}}$ are the weighting parameters to control the influence of the flow and motion regularization loss respectively, which are set to $0.5$ and $1e-3$ respectively~\cite{Jaganathan2021DeepRegistration}.

\subsection{Self-Supervised Registration Framework}
\label{sec:ss_dirn}
\begin{figure}
    \centering
    \includegraphics[width=\linewidth]{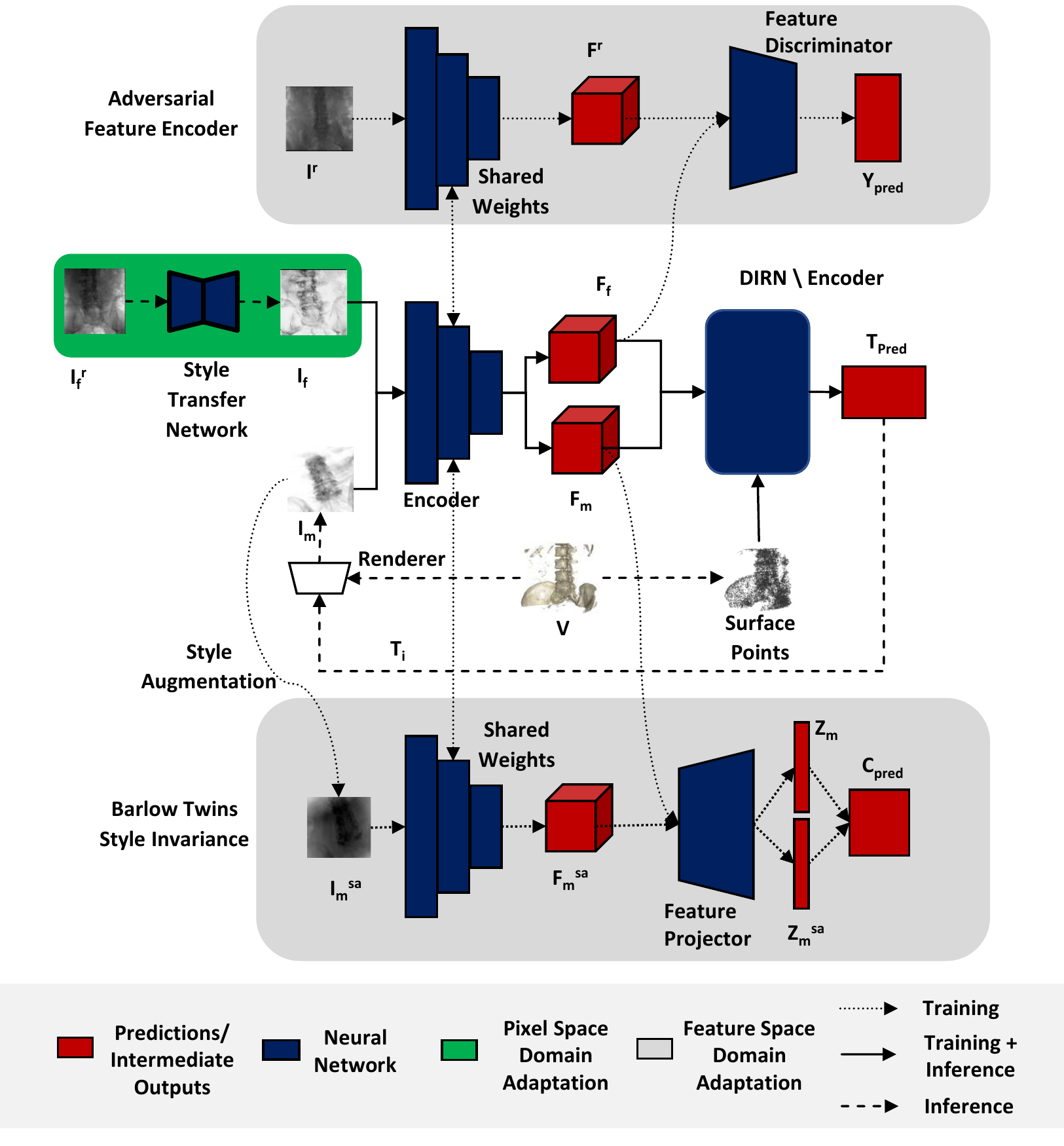}
    \caption{Our proposed registration framework with unsupervised pixel and feature space domain adaptation. We depict the encoder separately from the other DIRN modules for ease of visualization.}
    \label{fig:ss_dirn}
\end{figure}

In our self-supervised 2D/3D registration framework depicted in Figure~\ref{fig:ss_dirn}, we replace the use of paired training data ($\mathbf{I}_{f}^r, \mathbf{I}_{m} = \mathcal{R}(V, \mathbf{T}_{init}), \hat{\mathbf{T}}$) of DIRN framework with simulated training data. The simulated data is obtained using the bone projection style DRR images (Figure~\ref{fig:sim_xproj}) $\mathbf{I}_{f} = \mathcal{R}(V, \mathbf{T}_i)$ (instead of $\mathbf{I}_f^r$), $\mathbf{I}_{m} = \mathcal{R}(V, \mathbf{T}_j)$ rendered from arbitrary viewing directions $\mathbf{T}_i$ and $\mathbf{T}_j$ respectively. The ground truth registration matrix $\hat{\mathbf{T}}$ is computed from the relative transformation between  $\mathbf{T}_i$ and $\mathbf{T}_j$. Additionally, style augmented versions $\mathbf{I}_{f}^{sa}$, $\mathbf{I}_{m}^{sa}$ of bone projection style DRR $\mathbf{I}_{f}$ and $\mathbf{I}_{m}$ respectively are used for domain randomization. The style augmented version includes realistic projection style DRR and random style augmentations applied to DRR projections during training (Figure~\ref{fig:sim_xproj}). We henceforth refer to the network trained with simulated data including domain randomization as simulated DIRN. The inference still needs to be performed on the real X-ray images $\mathbf{I}_{f}^r$. The simplifying assumptions used to render DRRs compared to real X-ray images~\cite{Unberath2018}, lead to variations in image properties between the DRRs and the real X-ray images. To ensure that the simulated DIRN can perform with minimal performance gap on  $\mathbf{I}_{f}^r$, we perform unsupervised feature and pixel space domain adaptation. In feature space adaptation (Section~\ref{ss:feature_space_da}), our goal is to minimize the distribution shift between the encoded feature maps of $\mathbf{I}_{f}^r$ and  $\mathbf{I}_{f}$, while the features are optimal for DIRN. We perform the pixel space adaptation (Section~\ref{ss:pixel_space_adapt}) using unsupervised \mbox{image-to-image} style transfer network, where we transfer the input X-ray image $\mathbf{I}_{f}^r$ to the fixed bone projection style DRR $\mathbf{I}_{f}$. The training of the style transfer network is performed separately and coupled with feature adapted network during inference.

\subsubsection{Feature Space Domain Adaptation}
\label{ss:feature_space_da}
In feature space domain adaptation, the goal is to ensure the encoded feature map is well adapted for different style variations to the input image that can be encountered during the inference. Our feature space domain adaptation consists of Adversarial Feature Encoder (AFE) and Barlow Twins (BT)~\cite{ZbontarBarlowReduction} module trained together with DIRN in an end-to-end manner. Figure~\ref{fig:ss_dirn} depicts both the feature space domain adaptation modules and how it is trained together with DIRN. In the following, we introduce the modules separately and describe how the end-to-end training is performed together with the simulated DIRN. We use the image encoder from the original RAFT architecture~\cite{Teed2020RAFT:Flow} as proposed in DIRN~\cite{Jaganathan2021DeepRegistration}. 

\paragraph{Adversarial Feature Encoder}
Adversarial feature adaptation as a standalone module has been previously proposed~\cite{Ganin2015Domain-AdversarialNetworks, hoffman2018cycada} with application in unsupervised domain adaptation for medical images~\cite{Karaoglu2021AdversarialEstimation}. We use the adversarial feature loss (Eq.~\ref{eq:afe_loss}) computed using the encoded feature maps $\mathbf{F}_f$ and $\mathbf{F}^r$ from the bone style DRR $\mathbf{I}_f$ and unpaired real X-ray image $\mathbf{I}^r$. The feature discriminator $D_f$ (based on patch GAN discriminator~\cite{isola2017image}) is used to distinguish between the feature representations of the real and simulated images. The encoder can be trained with adversarial training~\cite{goodfellow2014generative} which ensures that the encoded feature distribution produced from both the X-ray and DRR images matches closely for similar structural content in both image. 

\begin{dmath}
     \mathcal{L}_{\mathrm{afe}} = \mathbb{E}_{\mathbf{x} \sim \mathbf{I}^s} [log D_f(E(\mathbf{x}))] \\ + 
     \mathbb{E}_{\mathbf{x} \sim \mathbf{I}^{r} } [ log D_f(1-E(\mathbf{x}) ) ]
     \label{eq:afe_loss}
\end{dmath}

\paragraph{Barlow Twins for Style Invariance}

We use the Barlow Twins(BT)~\cite{ZbontarBarlowReduction} module which was originally proposed for representation learning with recent advancements showing improvements in multi-modal representation learning~\cite{bardes2021vicreg}. We propose to use the original BT~\cite{ZbontarBarlowReduction} for our feature adaptation imparting style invariance. The BT loss (Eq.~\ref{eq:bt_loss}) is computed using  bone projection style DRR $\mathbf{I}_m$ and style augmented version $\mathbf{I}_{m}^{sa}$ of it. We compute the encoded feature maps  $\mathbf{F}_m$, $\mathbf{F}_{m}^{sa}$ for both the images $\mathbf{I}_m$ and $\mathbf{I}_{m}^{sa}$ respectively. The encoded feature map is projected to a $z$-dim embedding vector $\mathbf{Z}_m$ and $\mathbf{Z}_{m}^{sa}$ using the feature projector as proposed in ~\cite{ZbontarBarlowReduction}. Cross-correlation $\mathbf{C}_{pred}$ is computed between the embedded feature vectors $\mathbf{Z}_m$ and $\mathbf{Z}_{m}^{sa}$. Barlow twins loss~\cite{ZbontarBarlowReduction} (Eq.~\ref{eq:bt_loss}) which consists of invariance term and redundancy reduction term is used for training the encoder making it learn a feature representation that is invariant to the input style.
\begin{dmath}
     \mathcal{L}_{\mathrm{bt}} = \sum_i (1- \mathbf{C}_{pred}^{ii} )^2 + w_{\mathrm{red}} \sum_i \sum_{ i \neq j} (\mathbf{C}_{pred}^{ij})^2,
     \label{eq:bt_loss}
\end{dmath}
~~where $ w_{\mathrm{red}}$  is the weighting factor for the redundancy loss term~\cite{ZbontarBarlowReduction} which is set to 0.005.

\paragraph{End-to-end training with simulated DIRN}
Unsupervised feature adaptation for real X-ray images and different input styles can be achieved using the AFE and BT modules respectively. However, the features are not constrained to be optimal for the registration task (simulated DIRN). We propose to constrain the solution space for feature adaptation by training all the three modules (simulated DIRN, AFE, and BT) together in an end-to-end manner using shared weights for the encoder between the modules. Since AFE and BT does not require any paired registration data, we can perform task-specific unsupervised feature space domain adaption. We describe the training strategy used to train our network with Eq.~\ref{eq:feature_adapt} in Section~\ref{sec:training}.

\begin{dmath}
    \mathcal{L} = \mathcal{L}_{\mathrm{dirn}} +  w_{\mathrm{afe}}   \mathcal{L}_{\mathrm{afe}} + w_{\mathrm{dirn}} \mathcal{L}_{\mathrm{bt}},
    \label{eq:feature_adapt}
\end{dmath}

 ~~where $w_{\mathrm{afe}}$ and $ w_{\mathrm{bt}}$ are the weighting factors for the AFE and BT modules which are set to 0.2 and 0.05 respectively.

\subsubsection{Pixel Space Domain Adaptation}
\label{ss:pixel_space_adapt}
\begin{figure}
    \centering
    \includegraphics[width=\linewidth]{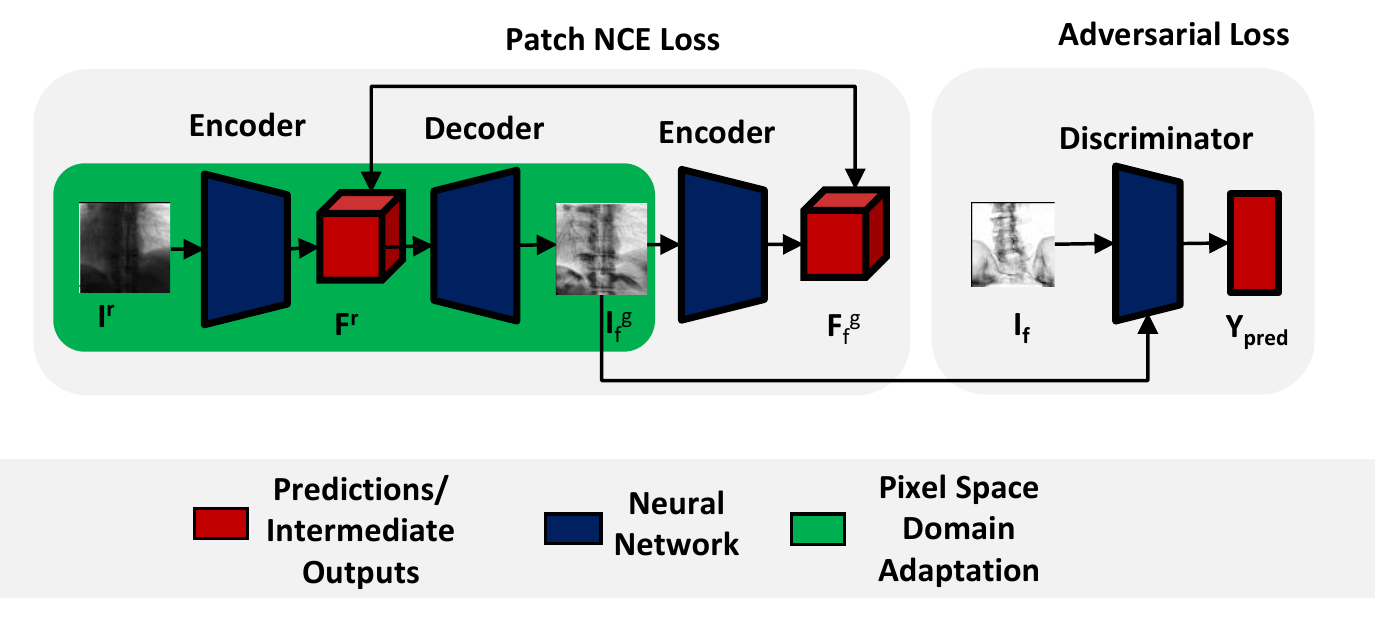}
    \caption{Unsupervised X-ray to DRR style transfer using Contrastive Unpaired Translation~\cite{Park2020ContrastiveTranslation}}
    \label{fig:cut}
\end{figure}

In pixel space adaptation, our goal is to make the pixel distribution of real X-ray images match to the simulated images encountered during training. We train our pixel space domain adaptation component separately, where we use an unsupervised image-to-image style transfer network. A common method to perform such style transfer is to use CycleGAN~\cite{Zhu2017UnpairedNetworks}. However, CycleGANs need to learn both forward and inverse mapping without any specific structural loss to preserve the structural content of the style transferred images. We instead use the recently proposed Contrastive Unpaired Translation (CUT)~\cite{Park2020ContrastiveTranslation} avoiding the need for learning forward and inverse mapping while also ensuring that the structural content is preserved with the patch NCE (PNCE) loss. We depict our X-ray to DRR style transfer using the CUT network in Figure~\ref{fig:cut}. The PNCE loss is computed using the input X-ray image and generated DRR on multiple feature levels of the encoder~\cite{Park2020ContrastiveTranslation}. A discriminator is used to distinguish between the real DRR images and the fake DRR images generated from input X-ray. 
The network is trained using Eq.~\ref{eq:cut_loss} with adversarial learning, where the goal of the generator is to produce DRR-styled images conditioned on the input X-ray images.
\begin{dmath}
    \mathcal{L}_{cut} = \mathcal{L}_{gan} +  \mathcal{L}_{pnce}  +\mathcal{L}^{id}_{pnce}
    \label{eq:cut_loss}
\end{dmath}
The $\mathcal{L}_{gan}$ is the Generative Adversarial Network (GAN) loss~\cite{goodfellow2014generative} for matching the distribution for the generated DRR with real DRR images. The $\mathcal{L}_{pnce}$ loss is the contrastive loss computed between the X-ray and generated fake DRR image patches ensuring the content of the style transferred image is preserved~\cite{Park2020ContrastiveTranslation}. The identity $\mathcal{L}_{pnce}$ is a learnable domain specific identity loss computed using the real DRR images~\cite{Park2020ContrastiveTranslation}. 
The CUT network is trained separately for X-ray to DRR transfer and is applied as a style transfer module during the inference to transfer the real X-ray images to DRR-styled images for our self-supervised 2D/3D registration network.

\subsection{Training and Inference}
\label{sec:training}
We pre-train simulated DIRN using the registration loss (Eq.~\ref{eq:DIRN}) similar to ~\cite{Jaganathan2021DeepRegistration}. We use simulated training with a fixed bone projection style DRR for 50 epochs and combine it with style augmented data for domain randomization for 50 epochs. We then fine-tune for feature space domain adaptation~\ref{sec:ss_dirn} using the combined loss (Eq.~\ref{eq:feature_adapt}) in an \mbox{end-to-end} manner for 20 epochs.
A cyclical learning rate between 1e-4 to 1e-6 is used during the simulated DIRN training phase. During fine-tuning for feature space adaptation a lower learning rate of 5e-6 is used for all DIRN modules except the image encoder. This is to make sure that our task performance is retained, while our encoder adapts for the different feature adaptations we perform using BT and AFE loss. The learning rate of other components (including the encoder) is set at 1e-4. We use a batch size of 16 for both pre-training and fine-tuning. We use Adam optimizer~\cite{kingma2014adam} to optimize all the network parameters. We update the feature discriminator of the AFE after each DIRN update. During inference except for the learned encoder weights, other feature adaptation modules are not used as depicted in Figure~\ref{fig:ss_dirn}.

We train our pixel space domain adaptation network separately on unpaired X-ray and DRR image datasets. We train the CUT network for 50 epochs using a learning rate at 1e-4 and batch size of 2 with Adam optimizer~\cite{kingma2014adam}. The trained generator network from CUT is then used during the inference as depicted in Figure~\ref{fig:ss_dirn}.
 We train our networks using PyTorch~\cite{paszke2017automatic}. Our models are trained using a single NVIDIA V100 Tesla GPU with 16GB memory. The inference is run on NVIDIA Titan X GPU with 12 GB memory with 10 iterations of DIRN for a registration sample.
Further implementation details are provided in the supplementary material.

\section{Experiments and Results}
\subsection{Experimental Setup}
\label{ss:exp_setup}
We describe the dataset used for training and evaluation in Section~\ref{ss:dataset} followed by the evaluation measures in Section~\ref{ss:evaluation_measures} and the baseline methods against which we benchmark in Section~\ref{ss:baseline}.

\subsubsection{Dataset}
\label{ss:dataset}
We use clinical Cone Beam CT (CBCT) reconstruction dataset consisting of the X-ray images with ground truth registration used for reconstruction along with the reconstructed 3D CBCT volumes. Our dataset is from the vertebra region, consisting of both thoracic and lumbar regions with 55 CBCT volumes acquired from 55 patients. The voxel spacing varies between 0.49 mm to 0.99 mm in all three dimensions. Due to the variations in the slice thickness, the number of X-ray images varies between 190 to 390 images per volume. The X-ray images have a resolution of $616\times480$ (width $\times$ height) with a pixel spacing of 0.616 mm.
We split our dataset into 43 patients for training, 6 patients for validation, and 6 patients for testing. We report all the results on the held-out test data set. 
A visualization of samples from our dataset is provided in the supplementary material.

In the case of supervised scenario (used for training supervised DIRN~\cite{Jaganathan2021DeepRegistration} for comparison), we create random initial transformations from the ground truth registration of the X-ray images with the initial registration error in the range of $[0, 30]$ mm. A training sample consists of $\mathbf{I}_{f}^r$, $\mathbf{T}_{init}$, $\hat{\mathbf{T}}$, $\mathbf{I}_m = \mathcal{R}(\mathbf{V},\mathbf{T}_{init})$ and $\mathbf{w}$.
In the self-supervised scenario, we retain the same random start position from $\hat{\mathbf{T}}$ of $\mathbf{I}_{f}^r$ and just replace it with $\mathbf{I}_{f}$, which can be rendered using $\hat{\mathbf{T}}$. We additionally have style augmented version of DRR images (Figure~\ref{fig:sim_xproj}) for domain randomization.
In total, we have 80,000 samples for training and validation for both supervised and self-supervised scenarios. To train the pixel space domain adaptation network and adversarial feature encoder we use a dataset of 16000 unpaired real X-ray and bone projection style DRR images from the vertebra region for the same set of patients.
Our test data set consists of 3600 samples from Anterior Posterior (AP) and lateral (LAT) views (as its the common views encountered during interventions~\cite{Liao2020, Miao2018}) for the 6 patients with initial registration error in the range of $[0,60]$ mm similar to ~\cite{Jaganathan2021DeepRegistration}. We use real X-ray images for all evaluations.

\subsubsection{Evaluation Measures}
\label{ss:evaluation_measures}
We use the standardized evaluation measure~\cite{kraats2004standardized} of mean Re-Projection Distance (mRPD) and the success ratio (SR) of mRPD $\leq$ 5.0 mm~\cite{otake2012automatic}. The mRPD indicates the overlay misalignment error and SR indicates the ratio of the number of samples to the total samples for which the mRPD $\leq$ 5.0 mm. The initial registration error is measured using the mean Target Registration Error (mTRE)~\cite{Schafferta, schaffert2020learning} indicates the 3D euclidean distance between the start position and the ground truth registration varies between $[0,60]$~mm.
\subsubsection{Baseline methods}
\label{ss:baseline}
We compare our proposed technique with both optimization-based and learning-based 2D/3D registration. For the optimization-based technique, we use DPPC~\cite{wang2020robust} as it showed state-of-the-art results compared to other optimization-based techniques~\cite{Wang2017,wang2020robust}. For learning-based 2D/3D registration, we directly compare with DIRN~\cite{Jaganathan2021DeepRegistration}, as we build our self-supervised registration framework on top of it. As an annotation-free baseline, we use the simulated DIRN with domain randomization~\cite{Grimm2021} which is commonly used in state-of-the-art registration networks that are trained only with simulated images~\cite{Grimm2021}.
We use the same training, validation, and test data for all the methods and the best hyper-parameter settings proposed in the respective original works.

\subsection{Results}
\label{ss:results}

We validate our unsupervised pixel space domain adaptation performance against standard supervised style transfer using pix2pix~\cite{isola2017image} in Section~\ref{ss:results_pixel_space}. Following, we present the ablation of the pixel and feature space domain adaptation components of our proposed network in Section~\ref{ss:results_ablation}. We then compare our method with other state-of-the-art techniques in Section~\ref{ss:results_sota}

\subsubsection{Pixel space domain adaptation}
\label{ss:results_pixel_space}

\begin{table}
\centering
\begin{tabular}{lll}
\toprule
 & mRPD [mm] $\downarrow$ & SR [\%]  $\uparrow$         \\
\midrule                            
Simulated  & 2.97 ± 0.99 & 66.2  \\
+ Unsupervised         & 2.35 ± 1.1        & 85.8           \\
+ Supervised   & \textbf{2.25 ± 1.0} & \textbf{86.8}       \\
\end{tabular}
\caption{Comparison of our unsupervised pixel space domain adaptation using CUT~\cite{Park2020ContrastiveTranslation} with supervised style transfer using pix2pix~\cite{isola2017image}. }
\label{tab:i2i_transfer}
\end{table}

\begin{figure}
    \centering
    \includegraphics[width=\linewidth]{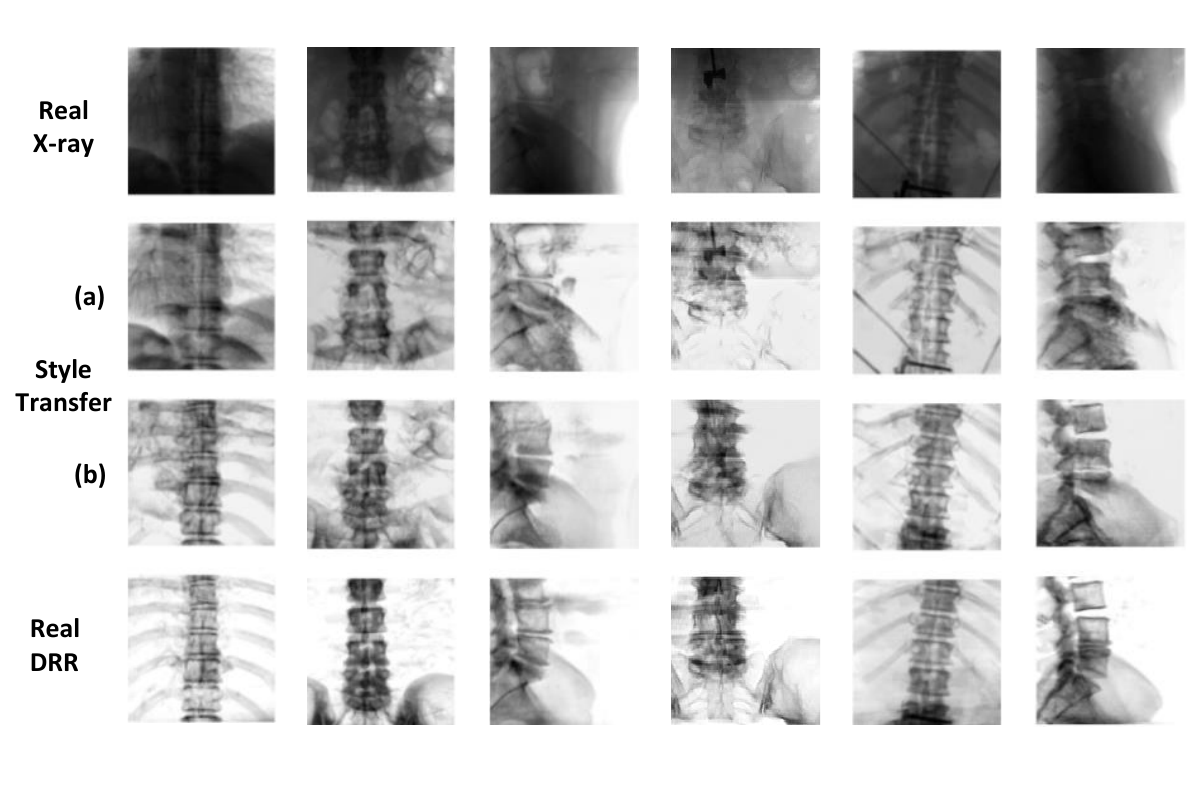}
    \caption{X-ray to DRR style transfer using (a) unsupervised CUT network (b) supervised pix2pix network along with real DRR. Each column indicates a test data sample.}
    \label{fig:i2i_transfer}
\end{figure}

We compare our pixel space adaptation using CUT network~\cite{Park2020ContrastiveTranslation} with supervised pix2pix network~\cite{isola2017image} for registration performance (Table~\ref{tab:i2i_transfer}) and X-ray to DRR style transfer (Figure~\ref{fig:i2i_transfer}). Our proposed unsupervised pixel space domain adaptation matches closely to the supervised style transfer network both in registration performance (Table~\ref{tab:i2i_transfer}) and the image appearance (Figure~\ref{fig:i2i_transfer}) indicating we are close to the maximum performance achievable for domain adaptation using style transfer.

\subsubsection{Ablation of domain adaptation components}
\label{ss:results_ablation}

\begin{table}
\centering
\begin{tabular}{lll} 
\toprule
 & mRPD [mm] $\downarrow$ & SR [\%]  $\uparrow$         \\
\midrule
Simulated           & 2.97
  ± 0.99       & 66.2           \\ 
+ Feature                   & 2.30
  ± 1.26       & 72.2           \\
+ Pixel                     & 2.35
  ± 1.1        & 85.8           \\ 
+ Feature + Pixel                        & \textbf{1.83 ± 1.16} & \textbf{90.1}   \\

\end{tabular}
\caption{Ablation of the different components of our self-supervised framework. }
\label{tab:ablation}
\end{table}
We compare the different components of our \mbox{self-supervised} 2D/3D registration framework and how each component drives the performance. In Table~\ref{tab:ablation}, we show the quantitative results comparing the different components of our registration framework. We start with our baseline scenario, where we use simulated training of DRR with domain randomization as proposed in~\cite{Grimm2021}. We add the feature space domain adaptation technique to train our network and this already shows a 6\% increase in SR compared to the simulated baseline. Adding the pixel space domain adaptation technique with the simulated baseline increases the SR by $19.6\%$. Our proposed self-supervised registration which consists of both pixel and feature space domain adaptation improves the SR by $23.9$\% from the simulated baseline. The registration error is also reduced from $2.97 \pm 0.99$ mm to $1.83 \pm 1.16$~mm.  Further analysis of simulated baseline and the visualization of registration error distribution is provided in the supplementary material.

\subsubsection{Comparison with state-of-the-art registration methods}
\label{ss:results_sota}
\begin{figure}
    \centering
    \includegraphics[width=\linewidth]{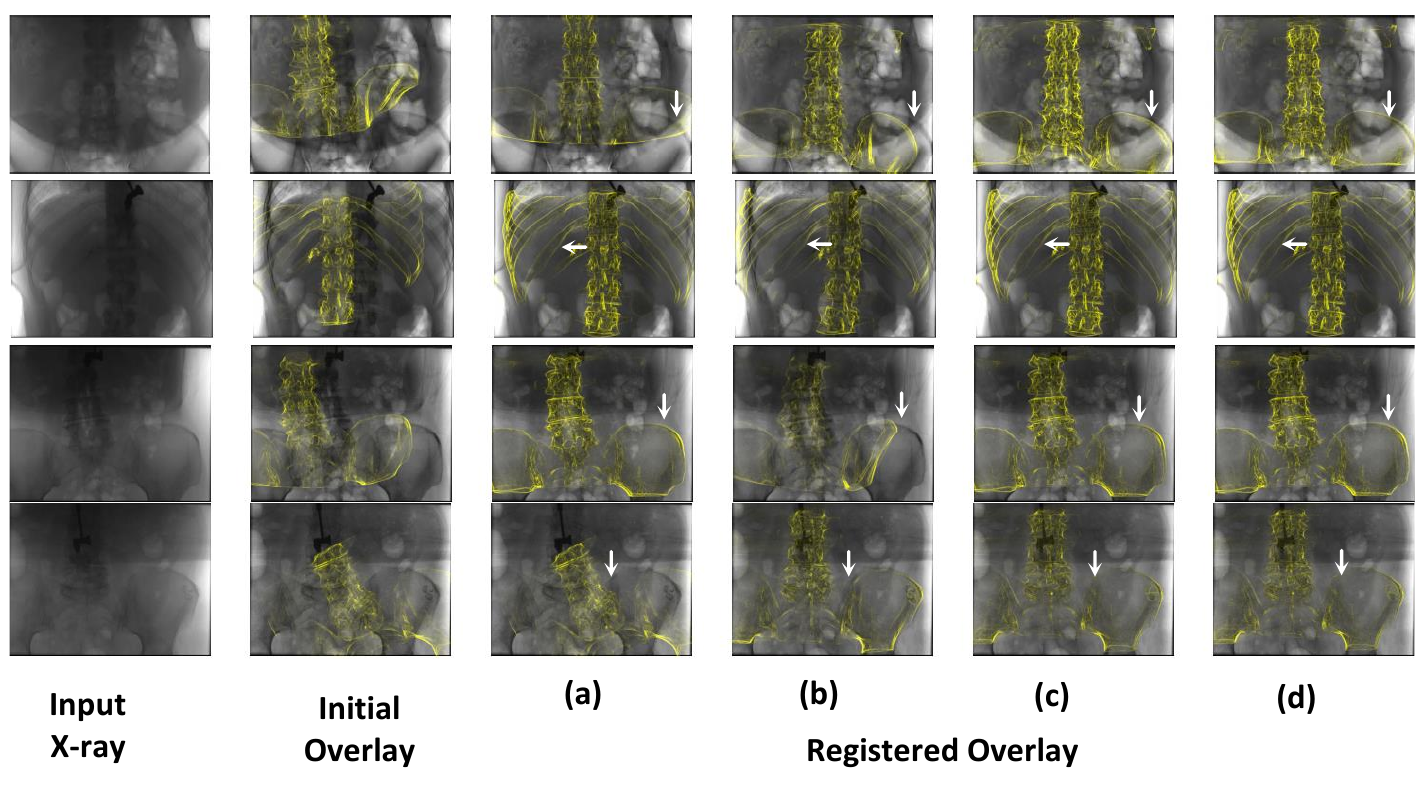}
    \caption{Visual comparison of overlays produced using  (a) optimization-based~\cite{Wang2017,wang2020robust} (b) simulated DIRN (includes domain randomization~\cite{Grimm2021}) (c) ours (d) supervised (DIRN)~\cite{Jaganathan2021DeepRegistration}. Each row indicates a test data sample.}
    \label{fig:overlay_compare}
\end{figure}

\begin{table}
\centering

\begin{tabular}{lll} 
\toprule
                        & mRPD [mm] $\downarrow$    & SR[\%] $\uparrow$   \\ 
\midrule
Optimization-based~\cite{Wang2017}           & \textbf{0.59 ± 0.26}       & 62.2   \\ 
Simulated           & 2.97 ± 0.99       & 66.2           \\ 
Ours           & 1.83 ± 1.16       & 90.1   \\ 
Supervised~\cite{Jaganathan2021DeepRegistration} & 0.65 ± 0.50  & \textbf{99.4}  \\

\end{tabular}
\caption{Comparison of our proposed method with other state-of-the-art techniques. }
\label{tab:sota}
\end{table}
We now compare with the different baseline methods which have shown state-of-the-art performance for different registration scenarios. Figure~\ref{fig:overlay_compare} shows the qualitative comparison of the overlays produced before and after registration. An arrow is marked on overlays produced using different methods to better illustrate the differences. The quantitative evaluation from Table~\ref{tab:sota} shows our proposed method achieves a SR of 90.1\% and has minimal performance drop compared to supervised method. The optimization-based technique lacks robustness to large initial misalignment resulting in lower SR compared to our proposed method. Figure~\ref{fig:overlay_compare} shows that overlays produced with our method is more accurate than simulated baseline and matches closely to the supervised DIRN~\cite{Jaganathan2021DeepRegistration}.

\section{Discussion}
Our novel unsupervised feature space domain adaptation couples Barlow Twins and Adversarial Feature Encoder, allowing us to increase the SR by 6\% compared to the simulated baseline without any additional parameters during inference (Table~\ref{tab:ablation}). Our unsupervised pixel space domain adaptation  using CUT network~\cite{Park2020ContrastiveTranslation} closely matches the performance of the supervised pix2pix-based style transfer network. The additional overhead is minimal for the style transfer network as it requires only a single forward pass of the generator during application. Also, unlike the CycleGAN~\cite{Zhu2017UnpairedNetworks}, we can directly learn the forward mapping between X-ray to DRR with structure consistency enforced by PatchNCE loss. Due to the strict design requirement of having structural consistency for style transfer, we skip the comparison with the CycleGAN as they do not satisfy it. CUT~\cite{Park2020ContrastiveTranslation} also outperforms it significantly for unsupervised image-to-image translation in~\cite{Park2020ContrastiveTranslation}.
We compare against the state-of-the-art 2D/3D registration techniques and clearly show that our proposed method achieves significant improvements to the SR compared to other annotation-free methods~\cite{wang2020robust} and closely matches the supervised learning-based technique~\cite{Jaganathan2021DeepRegistration}. Our proposed method also shows significant improvement in mRPD compared to the simulated baseline. However, one limitation of our proposed technique is the higher registration error compared to optimization-based and supervised registration methods.  Optimization-based refinement step~\cite{wang2020robust, Markelj2012, Miao2018, Liao2020} can be applied if lower registration error is necessary for the application.

\section{Conclusion}
~Our self-supervised framework enables training \mbox{learning-based} 2D/3D registration without the need for annotated paired datasets. We are one of the first to propose self-supervised 2D/3D registration targeted for dense correspondence-based 2D/3D registration networks~\cite{Jaganathan2021DeepRegistration, jaganathan2021learning, schaffert2020learning}, which are highly sensitive to the small changes in the image content. We combine the novel techniques from domain adaptation, self-supervised representation learning, and image-to-image translation to build a complete framework for self-supervised learning of dense \mbox{correspondence-based} 2D/3D registration networks. We achieve a high SR of 90.1\% on real X-ray images with a 23.9\% increase in SR compared to annotation-free alternatives. The mRPD is also reduced from $2.97 \pm 0.99$~mm for the simulated baseline to $1.83 \pm 1.16$~mm for our proposed method.

\clearpage

{\small
\bibliographystyle{ieee_fullname}
\bibliography{references}
}

\clearpage

\appendix
Our supplementary material provides further analysis of experiments in Appendix~\ref{sec:sup_analysis}, where we compare our simulated baseline without domain randomization. Additionally, we illustrate the domain gap between DRR and X-ray images, followed by visualizations of the registration error distribution for different variations of our proposed self-supervised framework. In Appendix~\ref{sec:sup_vis} we visualize additional samples comparing the overlays produced by the different state-of-the-art methods considered (Figure~\ref{fig:sup_additional_overlay}) and data samples from our clinical CBCT reconstruction dataset  (Figure~\ref{fig:dataset_overview}). We provide further implementation details in Appendix~\ref{sec:sup_imp_details}.

\section{Further Analysis of Experiments}
\label{sec:sup_analysis}

\subsection{Simulated Baseline}
We compare the simulated DIRN trained without domain randomization in Table~\ref{tab:sup_baseline_without_dr}, evaluated on real X-ray images of our test dataset. The SR drops from 66.2\% to 10\% for the network trained without domain randomization (using bone projection style DRR). Domain randomization significantly improves the performance on real X-ray images, as they have seen different styles during training. Thus, enabling us to have a strong baseline for the comparison with our proposed self-supervised framework.

\begin{table}[h!]
\centering
\begin{tabular}{lll} 
\toprule
 & mRPD [mm] $\downarrow$     & SR[\%] $\uparrow$    \\ 
\midrule
Simulated     & 2.97 ± 0.99 & 66.2   \\ 
 - DR  & 3.78 ± 0.83 & 10.0   \\ 
\end{tabular}
\caption{Comparison of Simulated DIRN with and without domain randomization evaluated on test dataset with real X-ray images. The simulated is our baseline which includes domain randomization and -DR indicates without domain randomization.}
\label{tab:sup_baseline_without_dr}
\end{table}

\subsection{DRR to X-ray Domain Gap}

\begin{table}[h!]
\centering
\begin{tabular}{lll} 
\toprule
 & mRPD [mm] $\downarrow$     & SR[\%] $\uparrow$    \\ 
\midrule
DRR Eval  &  0.27 ± 0.60 & 99.3        \\ 
X-ray Eval     & 2.97 ± 0.99 & 66.2   \\ 
\end{tabular}
\caption{Comparison of simulated DIRN (includes DR) evaluated on DRR (DRR Eval) and real X-ray images (X-ray Eval) from our test dataset for the same start positions.}
\label{tab:sup_baseline_comp_with_drr_eval}
\end{table}

\begin{figure}
    \centering
    \includegraphics[width=\linewidth]{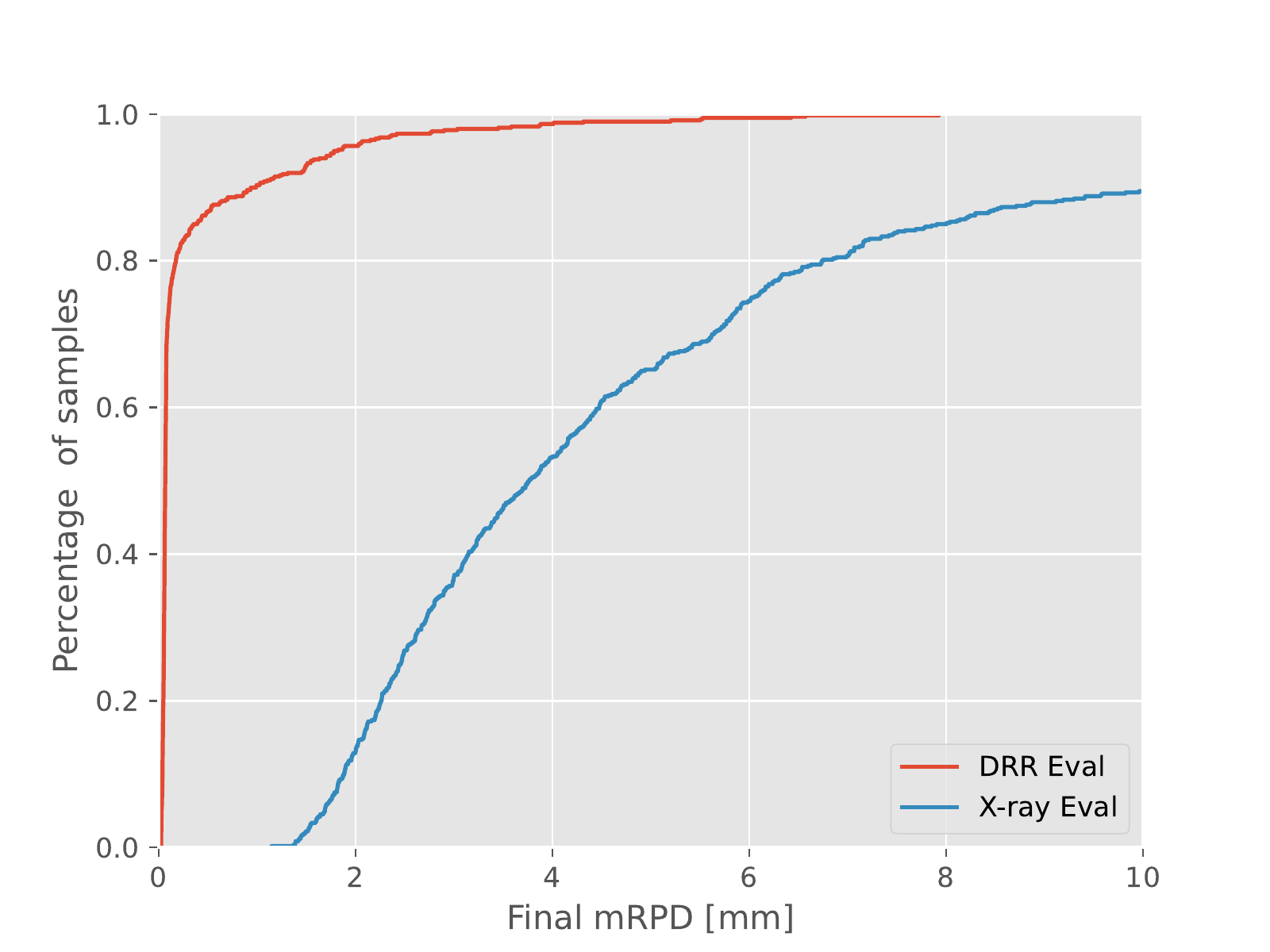}
    \caption{Comparison of final registration error using empirical cumulative distribution for DRR Eval and X-ray Eval of our simulated DIRN (includes DR), indicating the large domain gap that exists even after the application of domain randomization.}
    \label{fig:sup_cdf_xray_drr}
\end{figure}

We evaluated our simulated DIRN (includes DR) on the real X-ray and DRR images for the same start positions from our test dataset to illustrate the domain gap. As illustrated in Table~\ref{tab:sup_baseline_comp_with_drr_eval}, we achieve similar results to DIRN~\cite{Jaganathan2021DeepRegistration} when the source and target domain are same (DRR Eval). There is a huge drop in performance of our simulated DIRN  when evaluated on real X-ray images (X-ray Eval). In Figure~\ref{fig:sup_cdf_xray_drr}, we plot the cumulative registration error distribution for DRR and X-ray Eval of our simulated baseline network. The shift of the registration error towards higher values for X-ray Eval from the DRR Eval clearly illustrates the domain gap that exists even after the application of domain randomization.

\subsection{Ablation of Domain Adaptation Components}
\begin{figure}
    \centering
    \includegraphics[width=\linewidth]{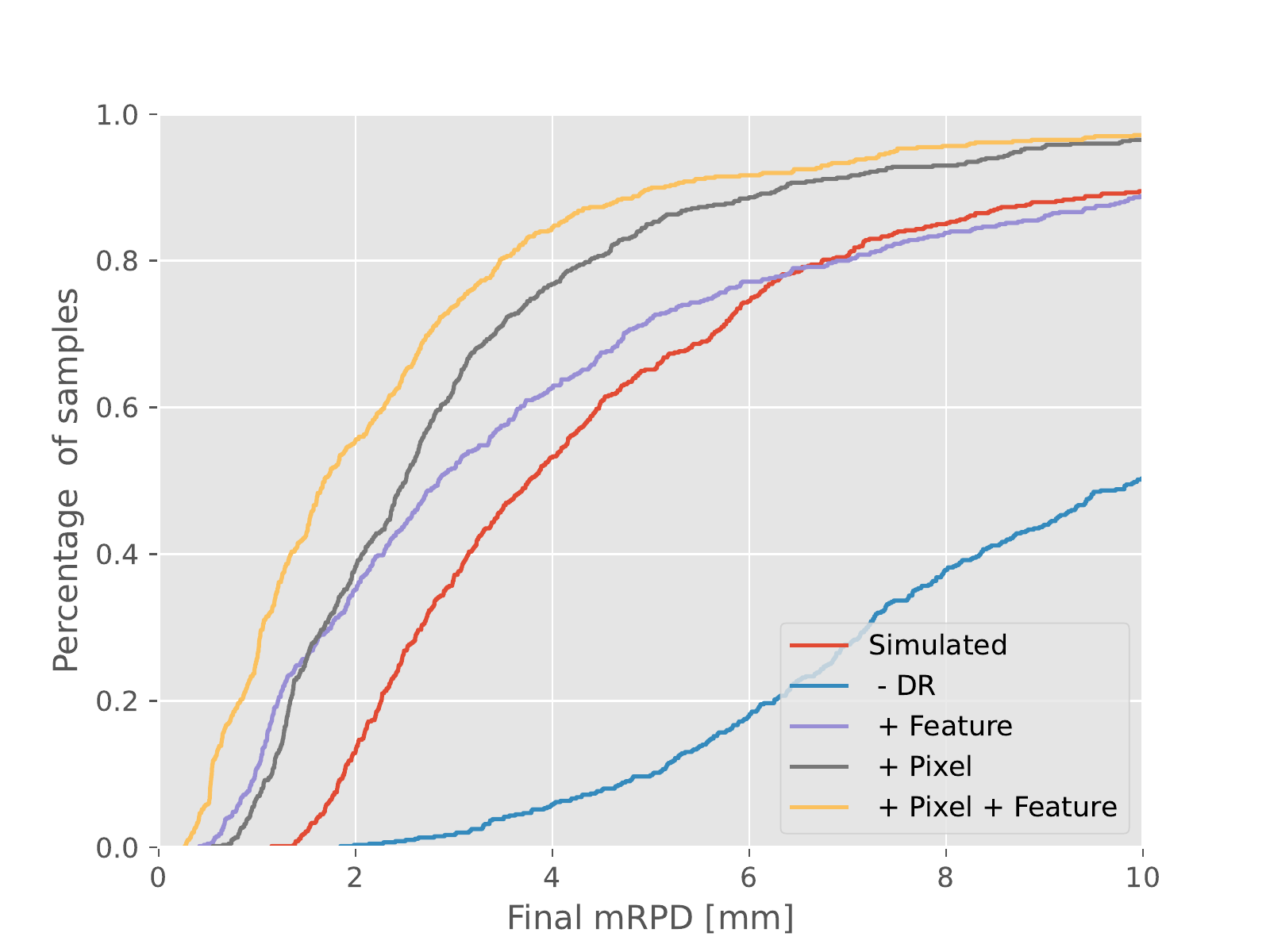}
    \caption{Comparison of final registration error using empirical cumulative distribution for different variations of our proposed framework.}
    \label{fig:sup_distribution_of_registration_error_cdf}
\end{figure}

\begin{figure}
    \centering
    \includegraphics[width=\linewidth]{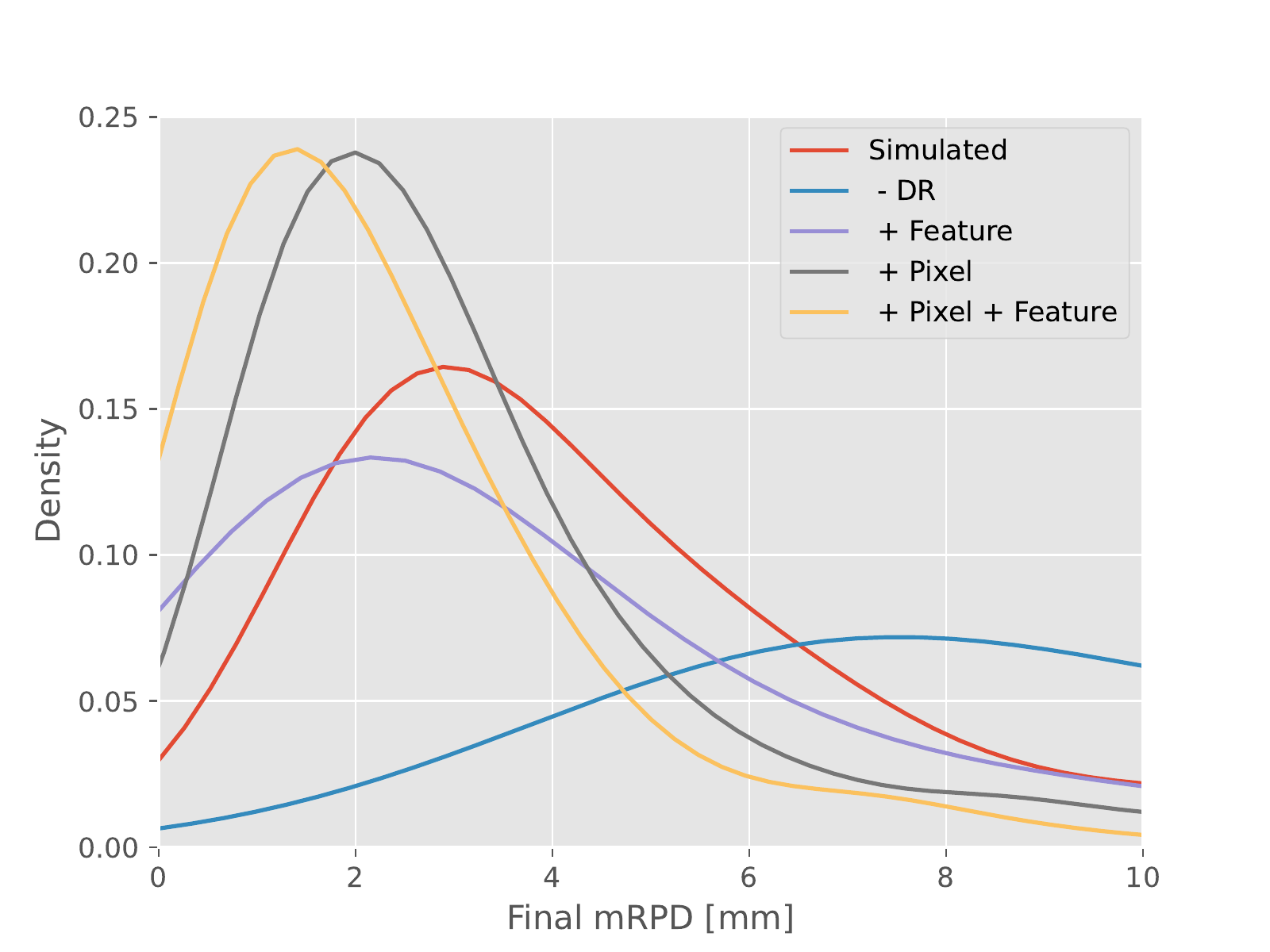}
    \caption{Comparison of final registration error using kernel density estimation for different variations of our proposed framework.}
    \label{fig:sup_distribution_of_registration_error}
\end{figure}

We visualize the cumulative distribution and kernel density distribution of the final registration error in Figure~\ref{fig:sup_distribution_of_registration_error_cdf} and Figure~\ref{fig:sup_distribution_of_registration_error} respectively for different variations of our framework.
Our proposed framework shows a significant shift to lower registration error compared to the simulated baseline. The standalone feature and pixel space additions also illustrate the performance gains of each component.

\section{Visualization}
\label{sec:sup_vis}
\subsection{Comparison of Registered Overlays}
\begin{figure*}[h!]
    \centering
    \includegraphics[width=\linewidth]{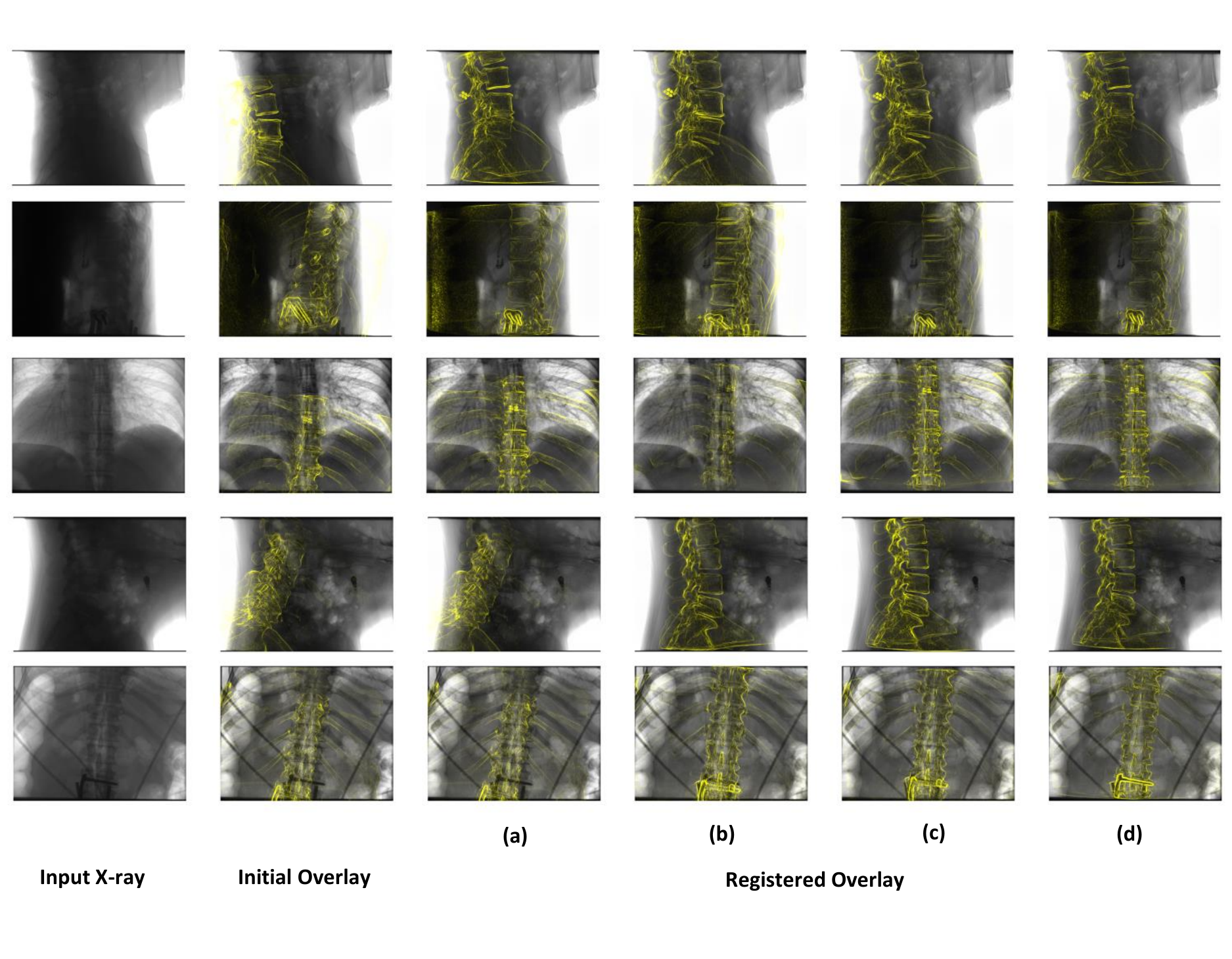}
    \caption{Additional samples from test dataset with comparison of overlays produced using (a) Optimization-based technique~\cite{wang2020robust}, (b) Simulated (with domain randomization~\cite{Grimm2021}), (c) our proposed method, and (d) supervised~\cite{Jaganathan2021DeepRegistration}. Each row represents a data sample from the test dataset.}
    \label{fig:sup_additional_overlay}
\end{figure*}

Figure~\ref{fig:sup_additional_overlay} shows additional examples from our test dataset, comparing the overlays produced. Each row depicts the comparison of the overlay produced from a single test sample for the different methods considered. 

\subsection{Dataset Visualization}
\begin{figure*}
    \centering
    \includegraphics[width=\linewidth]{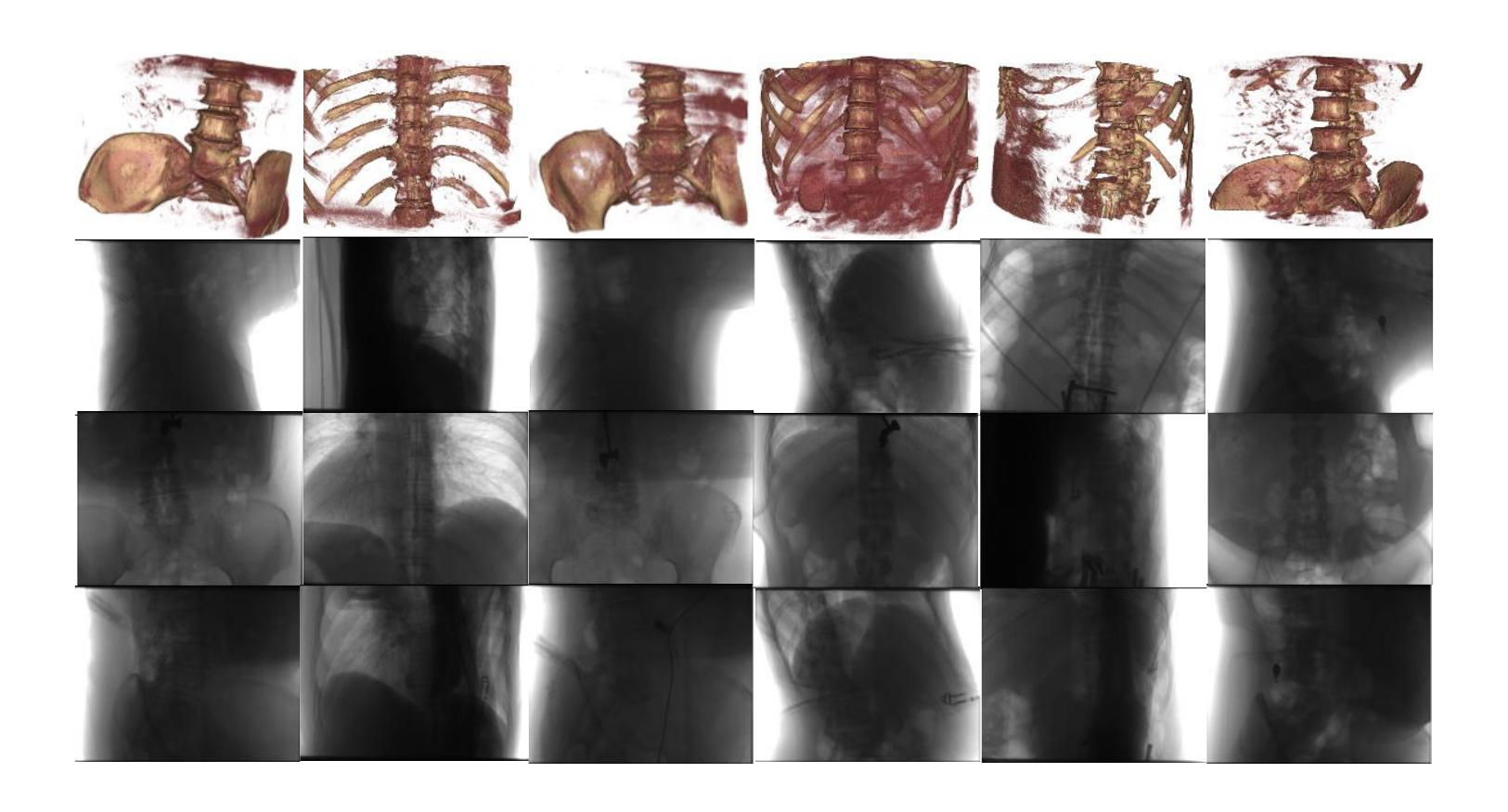}
    \caption{Exemplar data samples from our clinical CBCT dataset. The reconstructed volume is thresholded to better visualize the bone contours. Each column represents a reconstructed CBCT volume along with paired set of X-ray images used in reconstructing the CBCT volume.}
    \label{fig:dataset_overview}
\end{figure*}
Figure~\ref{fig:dataset_overview} shows example images from our clinical CBCT reconstruction dataset which includes the CBCT reconstructed volume along with the paired X-ray images.

\section{Implementation details}
\label{sec:sup_imp_details}
\subsection{Image Preprocessing}
The input images $\mathbf{I}$ (includes simulated $\mathbf{I}^{s}$ and real $\mathbf{I}^{r}$ X-ray images) are center cropped to a size of $480 \times 480$ from original image size of $640\times480$. The center cropped image is resized to $256\times256$ and fed as input to the networks. We normalize the pixel values using the dataset mean and standard deviation.

\subsection{Network Architecture Details}
Our self-supervised network consists of the registration network DIRN~\cite{Jaganathan2021DeepRegistration}, feature adaptation components (Adversarial Feature Encoders and Barlow Twins~\cite{ZbontarBarlowReduction}), and the unsupervised style transfer network~\cite{Park2020ContrastiveTranslation}. We use the architecture proposed in the respective original works, with the specific configuration used for our framework described below. The registration network DIRN~\cite{Jaganathan2021DeepRegistration} consists of RAFT~\cite{Teed2020RAFT:Flow} architecture for estimating the correspondence between the fixed $\mathbf{I}_f$ and moving $\mathbf{I}_m$ images. The RAFT architecture consists of a feature encoder and a context encoder. We input $\mathbf{I}_m$ to the context encoder and perform no domain adaptation since $\mathbf{I}_m$ is the fixed style bone projection DRR for both training and evaluation. We perform all the domain adaptations on the feature encoder as we would like to replace the simulated images $\mathbf{I}_f$ with the real X-ray images $\mathbf{I}_f^r$ during evaluation. Both the feature and context encoder are based on ResNet blocks~\cite{Teed2020RAFT:Flow}. The encoded feature map from the feature encoder is of the size $[256,32,32]$ for both $\mathbf{I}_m$ and $\mathbf{I}_f$. The RAFT uses iterative residual flow estimation for training and evaluation. We set the number of iterations for flow estimation to 6 for both training and evaluation. We use the PointNet++ architecture~\cite{qi2017pointnet++} for correspondence weighting as proposed in DIRN~\cite{Jaganathan2021DeepRegistration}. The single scale grouping-based segmentation architecture of PointNet++ which can output per-point classification is used. We replace the final layer with a Sigmoid activation function for predicting per-point weights in the range of $[0,1]$. The feature projector of the Barlow Twins consists of an MLP with three hidden layers of size $[512,256,128]$ that projects the encoded feature maps to 128-dimension embedding vector $\mathbf{Z}$. The feature discriminator of the adversarial feature encoder uses patch GAN~\cite{isola2017image} with a patch size of 8 and input channel dimension of 64. We use $1\times1$ convolution to match the encoded feature map to the input channel dimension of the patch GAN discriminator. The unsupervised style transfer network based on CUT~\cite{Park2020ContrastiveTranslation} uses a ResNet based generator consisting of 9 residual blocks~\cite{johnson2016perceptual} and patch GAN discriminator~\cite{isola2017image}, with a patch size of 16.

\end{document}